\documentclass[fleqn,usenatbib]{mnras}

\usepackage{newtxtext,newtxmath}

\usepackage[T1]{fontenc}
\usepackage{ae,aecompl}
\usepackage{booktabs}
\usepackage{soul}

\usepackage{enumitem}%
\usepackage{calc}
\usepackage{graphicx}	
\usepackage{amsmath}	
\usepackage[dvipsnames]{xcolor}
\usepackage{xcolor}
\usepackage{makecell}

\newcommand{\kms}{km\,s$^{-1}$}	
\definecolor{mypink}{rgb}{0.958, 0.188, 0.478}

\title[StarUnLink]{StarUnLink: identifying and mitigating signals from communications satellites in stellar spectral surveys}

\author[Bialek et al.]{
Spencer Bialek$^{1}$\thanks{E-mail: sbialek@uvic.ca.},
Sara Lucatello$^{2,3}$,
Sebastien Fabbro$^{1,4}$,
Kwang Moo Yi$^{5}$,
Kim A. Venn$^{1}$
\\$\;$
\\
$^{1}$ Department of Physics and Astronomy, University of Victoria, Victoria, BC, V8W 3P2, Canada\\
$^2$ INAF-Osservatorio Astronomico di Padova, Vicolo Osservatorio 5, 35122, Padova, Italy\\
$^3$ Institute for Advanced Studies, Technische Universität München, Lichtenbergstraße 2 a, 85748, Garching bei München, Germany \\
$^{4}$ National Research Council Herzberg Astronomy and Astrophysics, 
Victoria, BC, Canada\\
$^{5}$ Department of Computer Science, University of British Columbia, Vancouver, BC,  V6T 1Z4, Canada
Victoria, BC, Canada\\
}

\date{Accepted XXX. Received YYY; in original form ZZZ}

\pubyear{2023}

\begin{document}
\label{firstpage}
\pagerange{\pageref{firstpage}--\pageref{lastpage}}
\maketitle



\begin{abstract}
A relatively new concern for the forthcoming massive spectroscopic sky surveys is the impact of contamination from low earth orbit satellites. Several hundred thousand of these satellites are licensed for launch in the next few years and it has been estimated that, in some cases, up to a few percent of spectra could be contaminated when using wide field, multi-fiber spectrographs.
In this paper, a multi-staged approach is used to assess the practicality and limitations of identifying and minimizing the impact of satellite contamination in a WEAVE-like stellar spectral survey. We develop a series of convolutional-network based architectures to attempt identification, stellar parameter and chemical abundances recovery, and source separation of stellar spectra that we artificially contaminate with satellite (i.e. solar-like) spectra. Our results show that we are able to flag 67\% of all contaminated sources at a precision level of 80\% for low-resolution spectra and 96\% for high-resolution spectra. Additionally, we are able to remove the contamination from the spectra and recover the clean spectra with a $<$1\% reconstruction error. The errors in stellar parameter predictions reduce by up to a factor of 2-3 when either including contamination as an augmentation to a training set or by removing the contamination from the spectra, with overall better performance in the former case. 
The presented methods illustrate several machine learning mitigation strategies that can be implemented to improve stellar parameters for contaminated spectra in the WEAVE stellar spectroscopic survey and others like it.

\end{abstract}


\section{Introduction}

The number of satellites in low Earth orbit (LEO hereafter) has dramatically expanded since the turn of the century, increasing by at least an order of magnitude in the last decade. Commercial enterprises have become major players in this context, aiming at building infrastructure for global low-latency broadband Internet and other telecommunications needs. Several tens of thousands of LEO satellites from multiple companies have been licensed for launch by the end of the decade, and several hundred thousands more have been announced.

Besides potential implications for debris, there is a growing concern of
LEO satellites impacting our observations of astronomical sources in the night sky.
A recent study by \citet{Lawler2022} modeled the orbits of 65,000 artificial satellites (e.g. Starlink, OneWeb, Kuiper, and StarNet/GW) soon to be inhabiting our night sky and found that, at certain points throughout the night, there can be several hundred (or $\sim$7\% of the total visible point sources in the sky at $g<7$ within a couple of hours of sunrise or sunset) satellites bright enough to interfere with astronomical observations. 
The impact of LEO satellite on astronomy has been studied by several working groups supported by the National Science Foundation, American Astronomical Society, Canadian Astronomical Society, and the European Astronomical Society.
These studies have led to the
\textit{SATCON1} and 
\textit{SATCON2} reports \citep{SATCON1,SATCON2},  the \textit{Report on Mega-Constellations to the Government of Canada and the
Canadian Space Agency} \citep{MegaCons}, and the \textit{Dark and Quiet Skies for
Science and Society} reports\footnote{The reports can be found at \href{https://www.unoosa.org/oosa/en/ourwork/psa/schedule/2021/2021_dark_skies.html}{this website}}  \citep{walker2020,walker2021}

In the \textit{SATCON1} report, the impacts on the data collected at large spectroscopic facilities were noted to be larger than anticipated due to several factors: the size of the primary mirrors, the long integration times, and the brightness of a satellite when observing faint targets. 
In particular, the contamination from a bright satellite would likely go unnoticed until after the spectra are recorded, thus can only be detected in {\it post-processing}.
This could be a significant problem for the forthcoming sky surveys. 

Here we will focus on spectroscopic surveys that will make use of upcoming and planned wide-field, high multiplexing spectrographs, because of their requirements for extreme precision and  accuracy in understanding the dynamics and formation history of the Milky Way. In particular we will assess the impacts of LEO satellites on the European WEAVE Survey \citep{Dalton2018}, with implications that carry over to 4MOST \citep{dejong2019} (both are on 4m telescopes and are similarly impacted due to having similar characteristics such as diameter, average fiber density on the field-of-view, and spectral range/resolution) and to a lesser extent e.g. the 8.2m Subaru Telescope's Prime Focus Spectrograph \citep{Tamura2018}, and the 11.25m Maunakea Spectroscopic Explorer \citep[e.g.,][]{McConnachie2016}. 

The expected impact on spectroscopic observations is difficult to quantify, but \cite{hainaut2020} estimate that, for wide field, multi-fiber spectrographs, a few percent of spectra collected during the first and last hours of the night will be impacted by LEO satellite contamination, while \cite{bassa2022} estimate that up to 0.8\% of 4MOST spectra will be affected -- though we note these estimates were based on several tens of thousands of LEO satellites which, as previously mentioned, could be an underestimate for when 4MOST and WEAVE are collecting most of their data, and thus are likely lower limits. These surveys are designed to observe thousands of spectra per setting over a large field of view, with typically 10+ settings per night.
Over the lifetime of these surveys ($\sim$10 years), then 0.8\% per setting x 1000 stars per setting x 10 settings per night x 300 nights per year x 10 years = 240,000 contaminated spectra (out of 30 million spectra in the total survey). 


Because satellites will be abundant, physical mitigation strategies such as avoiding observations cannot be an option -- they will sacrifice too much in terms of observing time \citep{hu2022satellite}. Thus, alternative strategies are necessary, which is the aim of this paper. As the light from a bright satellite would primarily be reflected sunlight, then it is easy to simulate a contaminated observation of a stellar target by modelling it as an additive mix of stellar and solar-like spectra.
If there are solar spectra of varying S/N observed with the same instrument collecting stellar spectra, it is possible to create a large data set of artificially contaminated spectra. One can therefore attempt post-processing algorithms to identify which features of the spectra correspond to either solar or stellar light, and thereafter used for contamination mitigation strategies.

In this paper, we study the practicality of detecting and mitigating satellite contamination using a convolutional neural network and U-Net on \textit{post-processed} spectra. 
In Section 2, 
we clarify how we use the Gaia-ESO FLAMES-UVES spectra as an appropriate \textit{post-processed} test dataset.
In Section 3, we describe the ML methods used here. 
In Section 4, we apply our ML methods to contaminated Gaia-ESO FLAMES-UVES spectra to detect and remove the contamination and mitigate the errors in stellar parameter estimation.
We discuss our results in Section 5, including the limitations and best strategies going forward, and further applications of our method for bright sky subtraction and detecting spectroscopic binary stars more generally.

\section{Spectral Data
\label{sec:datasets}}




A set of 3100 R$\sim$47,000 FLAMES-UVES spectra \citep{pasquini2002installation} from the Gaia-ESO Survey (GES) DR4 \citep{sacco2014gaia}, in the wavelength range $483-530$ nm and with no continuum normalization applied, is used in this study. The GES targeted dwarfs and giants with metallicities between -2.5 and +0.5 dex and spectral types from O to M, so our sample contains a large variety of stars. These spectra were chosen as they cover a broad spectral region at optical-IR wavelengths, with high SNR, and high resolution.
To match the resolution and wavelength sampling of the WEAVE stellar survey, they were degraded and re-sampled to R=20,000 for the high-resolution (HR) mode, and R=5000 for the low-resolution (LR) mode, which we respectively refer to as the \texttt{WEAVE-GES-HR} sample and \texttt{WEAVE-GES-LR} sample hereafter. 
Note that this spectral range corresponds to the `green' arm of the WEAVE HR mode and a portion of the `blue' arm of the WEAVE LR mode; the red ($595-685$ nm) and blue ($404-465$ nm) arms of the HR mode and the red ($579-959$ nm) and entire blue ($366-606$ nm) arms of the LR mode will be used in future studies.

Additionally, the spectra are supplemented with a data set of solar/sky spectra\footnote{A set of 232 FLAMES-UVES solar spectra were collected from \href{https://www.eso.org/observing/dfo/quality/UVES/pipeline/FLAMES_solar_spectrum.html}{this website}}. 
A communications satellite is visible in the sky only because it is reflecting sunlight, thus a collection of solar spectra is crucial for simulating what a contaminated observation would look like. The details of the simulated contamination will be explained in Section~\ref{section:methods}.


\section{Machine Learning Methods}
\label{section:methods}


The application of ML
methods for analyzing astrophysical datasets has increased in popularity, particularly as computing power and complex ML algorithms become more powerful and accessible \citep{fluke2020surveying}.
Large observational surveys, as well as simulations, have provided massive datasets for developing ML tools with astrophysical applications. Large datasets are useful because they help to identify patterns and relationships in the data that might not be apparent with smaller datasets. With more data of high quality, models can learn to generalize better due to the diversity of examples, meaning they can make accurate and robust predictions about new, unseen data.

While other tools could be used to identify solar contamination, e.g. a traditional matched filter with solar spectrum template, we decided to use ML because of its computational efficiency (a dataset of millions of spectra could be analyzed in a couple of minutes) and history of successful implementations in astronomy (see \citealt{sen2022astronomical} for a recent review).

A neural network (NN) is a function that transforms an input into a desired output. The function is composed of many parameters, arranged in layers with non-linear activation functions, which form a highly non-linear combination of the input features, and allows for complex mappings to be represented with high precision.  
A convolutional neural network (CNN), in which a sequence of learned filters followed by a series of learned inter-connected nodes, can transform a stellar spectrum into a prediction of associated stellar parameters.
NNs and CNNs have proven to be effective at deconstructing and modelling the components of stellar spectra \citep[e.g.,][]{fabbro2018, leung2019, ting2019, guiglion2020radial, obriain2020, bialek2020, sharma2020application, zhao2022automated}.

In this section, we describe the ML methods we have built to identify and demix contaminated spectra in the \textit{post-processed} spectra in the Gaia-ESO FLAMES-UVES survey.

\subsection{Creating the training and test sets}

The sample of 3100 FLAMES-UVES spectra was split 80\%/20\% into reference/test sets, with a further 85\%/15\% split of the reference set to acquire training/validation sets. 
Within each set, a random stellar spectrum was chosen, contaminated to a random fraction (between 0-50\% of the median intensity value) with a randomly chosen solar spectrum from a set of 232 solar spectra, degraded from a resolution of 47,000 to 20,000 for \texttt{WEAVE-GES-HR} and to 5000 for \texttt{WEAVE-GES-LR}, and finally re-binned to the respective HR or LR WEAVE wavelength grid. Spectra in the reference set were additionally uniformly radial velocity shifted (|$v_{rad} | < 200$km/s) before the contamination was applied. The boundaries on the contamination were chosen based on Figure 6 from \cite{bassa2022} which shows the effective magnitude of LEO satellites to be around 21 -- given the normal exposure times expected for WEAVE ($\sim$600-1200s) -- and on the expected limiting magnitude of WEAVE to be G=20.7 in the low-resolution mode \citep{jin2022wide}. Note the expected limiting magnitude for the high-resolution spectra is G=16 so the contamination problem is far more problematic in the low-resolution spectra.
The contaminated spectrum was obtained with the formula 
\[\mathbf{x}_{\mathrm{contaminated}} = \mathbf{x}_{\mathrm{stellar}} + c \times \frac{\tilde{\mathbf{x}}_{\mathrm{stellar}}\mathbf{x}_{\mathrm{solar}}}{\tilde{\mathbf{x}}_{\mathrm{solar}}}\] 
where $\mathbf{x}_{\mathrm{stellar}}$ is a radial velocity shifted stellar spectrum, $\mathbf{x}_{\mathrm{solar}}$ is a solar spectrum, $c$ is the contamination fraction, and $\tilde{\mathbf{x}}$ refers to the median. The process was repeated until the final training/validation/test sets contained 35000/5000/10000 spectra (the performance did not increase when training on more samples). All data sets retained the original, clean spectra, such that e.g. the training set included 35000 contaminated and their corresponding 35000 clean spectra, important for corrections when imbalanced training was taking place.

\subsection{Neural networks}

\subsubsection{Classification network for detection}

A standard CNN was implemented for the classification task (i.e., our StarNet architecture, \citealt{fabbro2018, bialek2020}).
The input was an individual stellar spectrum and the output prediction was a score between 0 (clean) and 1 (contaminated). 
A binary cross-entropy (BCE) loss function was used to compare the CNN's output prediction with the target label of the input spectrum, such that the output of the trained CNN can be thought of as a prediction of the probability that the spectrum is contaminated (though it is not guaranteed to be a probability - i.e. calibrated). The Adam algorithm \citep{kingma2014adam} was used for the gradient-based optimization of the BCE, with a learning rate that was decreased when the performance on the validation set plateaued. The model that scored the lowest BCE on the validation set was saved, to ensure the final model did not overfit to the training data, and used for all tests. 

\subsubsection{Regression network for stellar parameters}

The same CNN and training scheme was used as in the classification task; however, the last layer was changed to linear outputs, and the loss function was changed to the mean-squared error (MSE).
These were then used to regress on the stellar parameters of the individual stellar spectra. The mean and standard deviation of the distributions of stellar parameters were used to normalize the labels (for zero mean and unit variance), and then to de-normalize the inferred stellar parameters.

\subsubsection{U-Net for removing contamination}
\label{section: waveunet}

\begin{figure}
    \centering
    \includegraphics[width=0.5\textwidth]{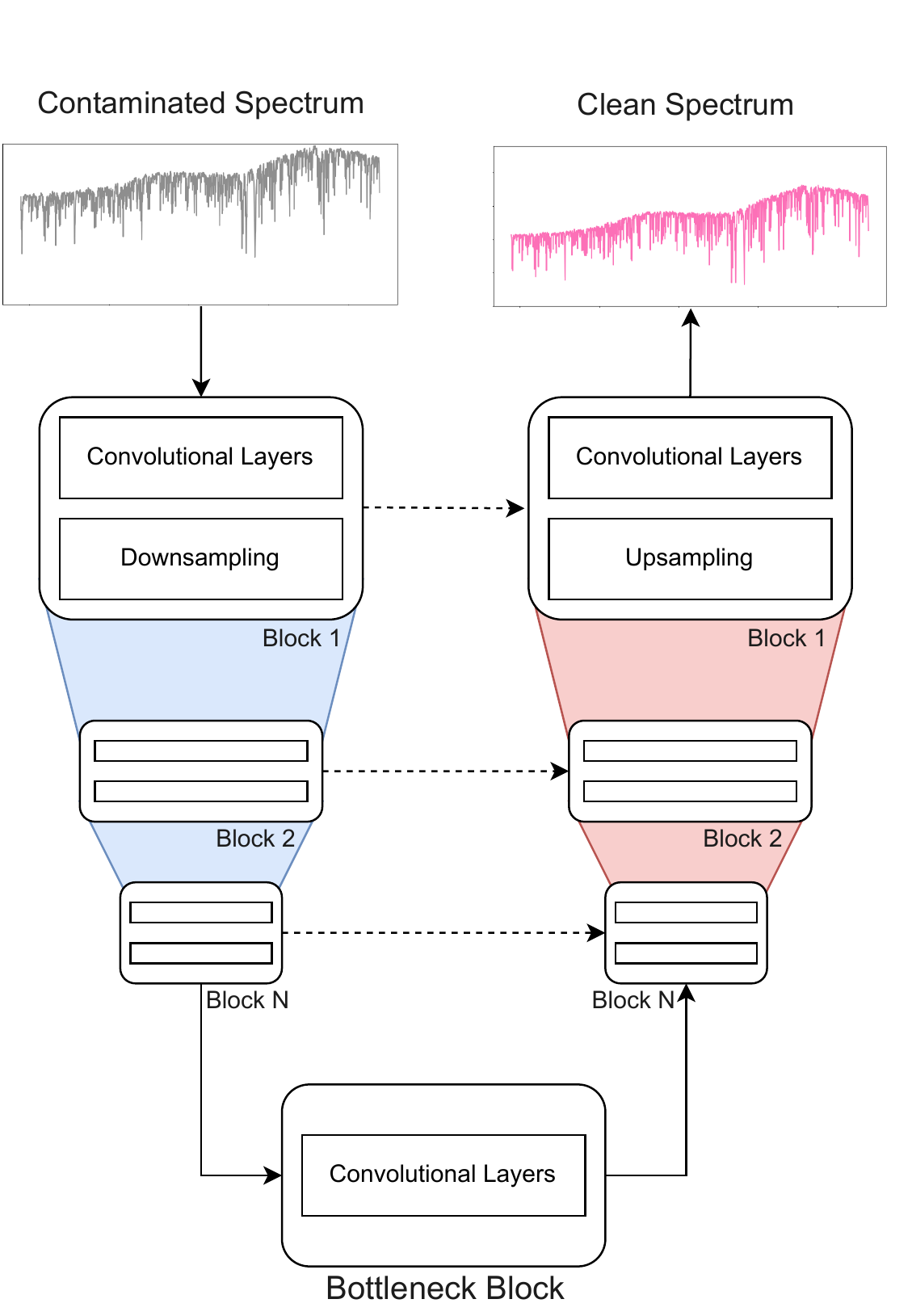}
    \caption{A schematic illustration of the Wave U-Net removing satellite contamination from a stellar spectrum. The details of each block can be found in the project's  \href{https://github.com/Spiffical/starlink}{Github repository}.}
    \label{fig:waveunet}
\end{figure}

U-Nets are CNNs that were originally created for biomedical imaging segmentation tasks \citep{ronneberger2015u}. These include downsampling and upsampling layers that encode then decode data, and they combine high-resolution information from downsampled layers with subsequent upsampling layers via skip connections which helps to prevent information loss and vanishing gradients.

After a few trials on U-Net architectures, a particular adaptation of the U-Net to one-dimensional domains, \textit{Wave U-Net} \citep{stoller2018wave}, was adopted as a base architecture. Wave U-Net was created for audio source separation and introduced architectural improvements such as an upsampling technique and context-aware prediction framework to reduce artifacts in its output.  Empirical results of Wave U-Net showed its capacity in isolating raw wave forms from various instruments (e.g. bass, guitar, drums, voice). In a similar manner, Wave U-Net can be given a raw spectrum with LEO satellite contamination and tasked with inferring -- and therefore isolating -- the pure stellar spectrum. Figure \ref{fig:waveunet} shows schematically how Wave U-Net works on a stellar spectrum. We note an improved version of Wave U-Net was used in this study\footnote{details of the architecture improvements can be found at \href{https://github.com/f90/Wave-U-Net-Pytorch}{https://github.com/f90/Wave-U-Net-Pytorch}}.


A similar training scheme as the previous methods was used but a mean absolute error (MAE) regression loss was used instead of an MSE loss because the MSE led to an undesired amount of smoothing in the reconstructed spectra. 

\subsubsection{An ensemble of models}

If an ensemble of NNs with random initializations are independently trained on the same data and given the same task, then every NN will settle to a different minimum of the loss function owing to the stochastic nature of the training procedure. 
As such, each NN will have learned something unique about the training data, and no single NN in the ensemble can necessarily be considered the best. It is therefore an advantage to utilize each NN in the ensemble to make the most robust predictions. Following our similar procedure as in \cite{bialek2020}, an ensemble of CNNs was trained for the detection and 
probability of an individual spectrum being contaminated.

\section{Results}
\label{section:contamination-results}

The \texttt{WEAVE-GES-HR} and \texttt{WEAVE-GES-LR} spectra were used to assess the performance of the CNN and Wave U-Net and showed the following results.

\subsection{Detection}

\begin{figure}
    \centering
    \includegraphics[width=0.5\textwidth]{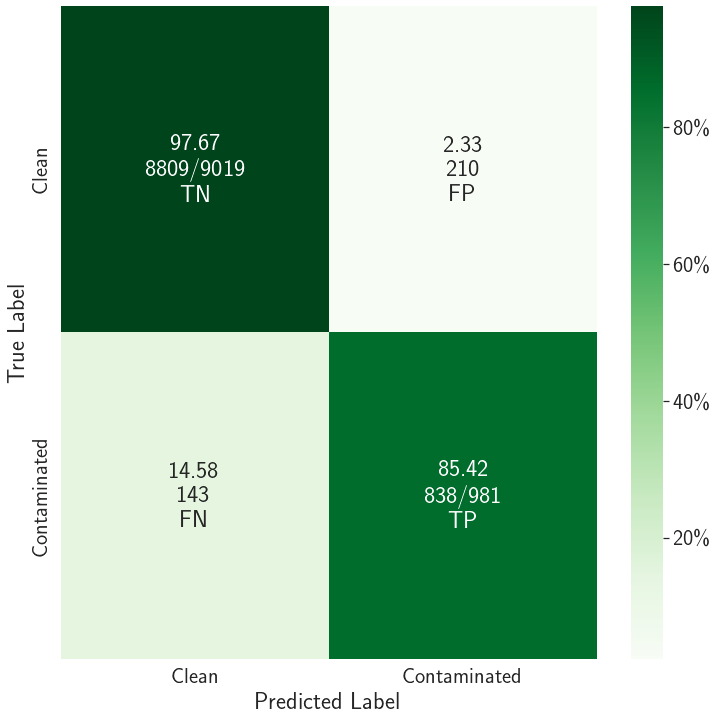}
    \caption{A confusion matrix showing information about the rates of true positives, true negatives, false positives, and false negatives of the \texttt{WEAVE-GES-LR} sample. The test set of 10,000 spectra was divided such that 10\% were contaminated spectra and 90\% were uncontaminated (clean) spectra. The model had a true positive detection rate of 85.42\% and false positive detection rate of 2.33\%. A decision boundary of 0.7 was used for this plot; the false positive rate can be decreased by raising the decision boundary (i.e. accepting positive predictions only when the model is very confident) at the cost of decreasing the true positive rate. Note: similar results were found with the \texttt{WEAVE-GES-HR} sample.}
    \label{fig:confusion_matrix}
\end{figure}

For the detection task, a spectrum was considered to be contaminated if its contamination level was $>$1\%. In the test set of 10,000 spectra, 1,000 were contaminated and 9,000 were clean, uncontaminated spectra. Figure \ref{fig:confusion_matrix} shows a confusion matrix for the contamination detection on the \texttt{WEAVE-GES-LR} test set with the decision threshold set to 0.7, i.e. spectra were considered contaminated if the output prediction of the CNN was $>$0.7. The true negative detection rate was an impressive 97.67\%, meaning that, in this particular case, only 2.33\% of clean spectra (210 total) were misclassified. Meanwhile, the true positive detection rate was 85.42\% such that 14.58\% of contaminated spectra (143 total) were misclassified as clean spectra. The results on the \texttt{WEAVE-GES-HR} sample were similar. Depending on the decision threshold chosen, these variables can change based on the desired optimization; the next section explores this.

\subsubsection{Fine-tuning the decision threshold}

Since the CNN outputs a prediction between 0 and 1, there must be a choice made about when to consider a prediction to be positive or negative. A natural choice for this \textit{decision threshold} might be 0.5 if calibrated, i.e. predictions above and below this value are the contaminated and clean spectra, respectively. If the goal is to detect contaminated spectra, the threshold can be lowered to capture fainter contaminations, or raised for only the most confident predictions for contaminated spectra.

\begin{figure}
    \centering
    \includegraphics[width=0.5\textwidth]{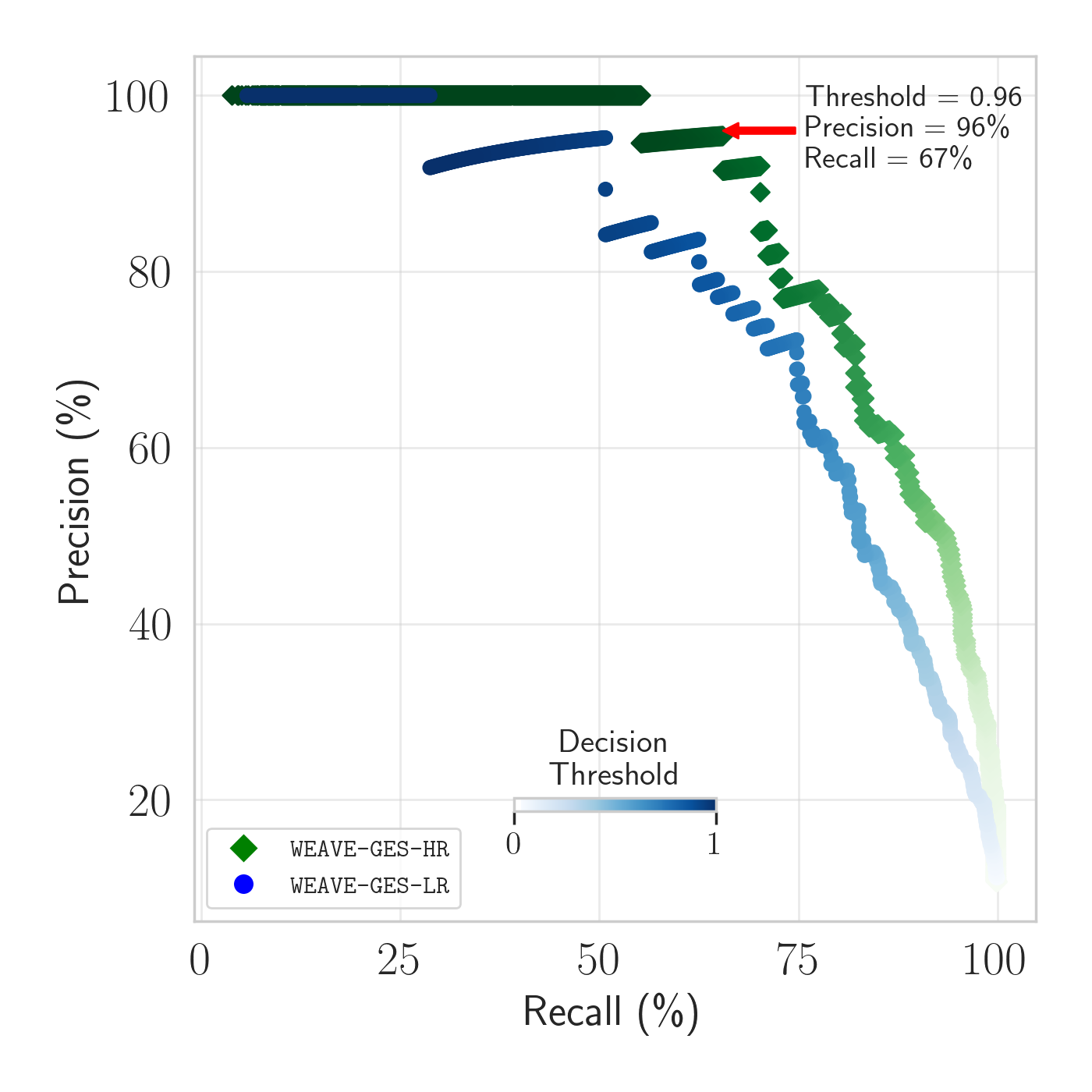}
    \caption{A precision-recall plot showing how effectively two CNNs separately trained to identify contaminated \texttt{WEAVE-GES-HR} and \texttt{WEAVE-GES-LR} spectra were able to identify contaminated spectra depending on the decision threshold chosen. As an example, for the \texttt{WEAVE-GES-HR} spectra, the threshold could be chosen such that the precision and recall of the identification procedure are 96\% and 67\% respectively, i.e. extremely precise yet missing 33\% of the true 
    contaminants.}
    \label{fig:precision-recall}
\end{figure}

To illustrate the prediction dynamics as the decision boundary is modified, a Precision-Recall analysis was conducted. Mathematically, precision and recall are defined as:
\[ Precision = \frac{TP}{TP+FP} \]
\[ Recall = \frac{TP}{TP+FN} \]

where TP is the true positives, FP is the false positives, and FN is the false negatives. Precision is a measure of the \textit{quality} of the CNN's predictions, whereas recall is a measure of the fraction of contaminated sources that were identified. Figure \ref{fig:precision-recall} shows how the precision and recall change as a function of decision threshold for the \texttt{WEAVE-GES-HR} and \texttt{WEAVE-GES-LR} samples: generally as the decision boundary increases, the precision increases and the recall decreases. The performance on the \texttt{WEAVE-GES-HR} is better overall, likely because the increased resolution provides more information to distinguish stellar features from solar features.

In the case of identifying contaminated spectra, it is probably desirable to optimize the decision threshold such that there is a high precision while minimally sacrificing the recall, e.g. the values can be optimized such that the model achieves 96\% precision and 67\% recall if a decision threshold of 0.96 is chosen: while the model misses 33\% of the total contaminated sources with this threshold, the trade-off is that there is trust when it does identify contamination. The precision drops to 80\% at the same recall level for the \texttt{WEAVE-GES-LR} sample.

\subsubsection{Undetected contamination}

To understand how the model was failing at properly identifying all of the contaminated sources, we examined how well the model recovered them depending on their contamination level and stellar parameters. 

\begin{figure}
    \centering
    \includegraphics[width=0.5\textwidth]{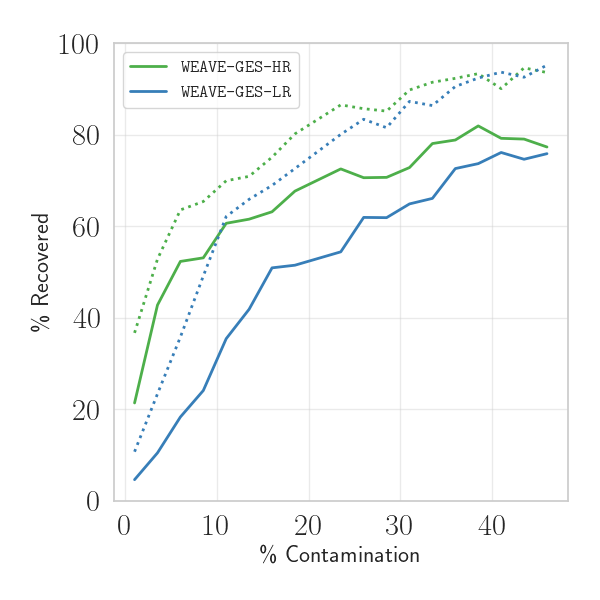}
    \caption{Two CNNs were separately trained to identify contaminated \texttt{WEAVE-GES-HR} and \texttt{WEAVE-GES-LR} spectra and the fraction of sources properly identified as being contaminated (i.e. the recall) was computed in bins of contamination level for decision thresholds of 0.95 (solid line) and 0.80 (dotted line). The majority of false negatives reside in the low ($<$5\%) contamination regime.}
    \label{fig:percent.recovered.vs.contam}
\end{figure}

\begin{figure*}
    \centering
    \includegraphics[width=0.9\textwidth]{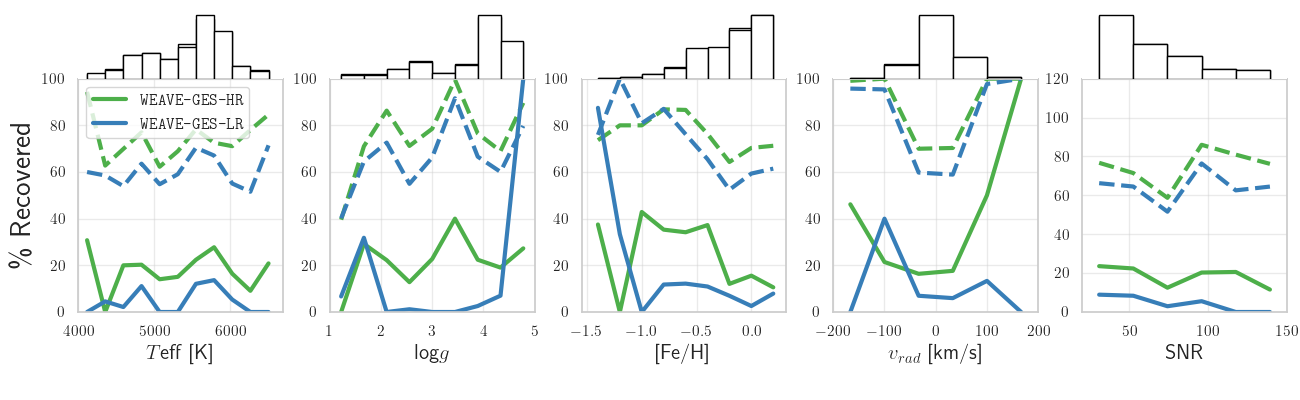}
    \caption{The recovery fraction (i.e. recall) of identified contaminated sources as a function of stellar parameter for the \texttt{WEAVE-GES-HR} and \texttt{WEAVE-GES-LR} samples, split into sources with high contamination (between 5\% and 50\%; dashed lines) and low contamination ($<5$\%; solid lines), using a decision threshold of 0.95. Also shown are the histograms of stellar parameters; low number statistics in certain parameter ranges cause spurious features.
    }
    \label{fig:precent.recovered.vs.stellar.params}
\end{figure*}

The recall as a function of contamination gives a more detailed understanding of the completeness of the model and is shown in Figure \ref{fig:percent.recovered.vs.contam} for two CNNs trained separately on the \texttt{WEAVE-GES-HR} and \texttt{WEAVE-GES-LR} samples.
Using a decision threshold of 0.95, the \texttt{WEAVE-GES-HR} CNN recovered approximately 20\% of contaminated sources at a level of 2-3\% contamination, but the recovery fraction increases rapidly to 60-80\% above 10\% contamination. When lowering the decision threshold to 0.80, the recovery rate approximately doubles in the low contamination regime and rises to 80-90\% for higher levels of contamination. The \texttt{WEAVE-GES-LR} CNN showed similar trends albeit with a decreased performance overall. The false negatives are concentrated in the low ($<$5\%) contamination regime.

To uncover any potential biases in detection with respect to the stellar parameters, the recall for the \texttt{WEAVE-GES-HR} and \texttt{WEAVE-GES-LR} CNNs was computed as a function of the parameters $T_{\textrm{eff}}$, log$g$, [Fe/H], $v_{\textrm{rad}}$, and SNR, using a decision threshold of 0.95, as seen in Figure \ref{fig:precent.recovered.vs.stellar.params}. One would expect the recovery fraction to decrease around solar values ($T_{\textrm{eff}}$ = 5777 K, log$g$ = 4.44, and [Fe/H] = 0, \citealt{de2014photometric}) since it would be difficult to distinguish a solar spectrum mixed with a solar-like spectrum. There indeed appears to be a local dip at solar log$g$ and a decreasing trend towards solar metallicity. The most obvious trend occurs with radial velocities: the recovery rate increases substantially at higher absolute velocities. Interestingly no clear trends are seen with SNR, perhaps because the SNR of each spectrum is sufficiently high.  Lastly, there are several anomalous or sporadic points where e.g. the recovery fraction is zero or nearly zero, but this is likely due to low numbers of samples in those regions (creating problems for both training the model and testing the model with small number statistics). 

Taken together, Figures \ref{fig:percent.recovered.vs.contam} and \ref{fig:precent.recovered.vs.stellar.params} imply that (1) a spectrum with solar-like surface gravity and metallicity presents an increased difficulty for the CNN classification scheme, (2) there is no obvious correlation between the recovery rate and effective temperature or SNR, and (3) the most significant factors in detectability are contamination levels and the relative difference between the radial velocity of the stellar and contaminating source. 

\subsection{Stellar parameters prediction}

The stellar parameters $T_{\textrm{eff}}$, log$g$, and [Fe/H], along with the absolute abundances A(Ca), A(Mg), A(O), A(S), and A(Ti), were predicted with a CNN. For the \texttt{WEAVE-GES-LR} sample, two models were trained: one on 35,000 spectra \textit{with} contamination between 0\% and 50\% (\texttt{CNN+}) and one on 35,000 spectra \textit{without} contamination (\texttt{CNN-}), and were used in the following tests. Note: the results on \texttt{WEAVE-GES-HR} can be seen in Appendix \ref{appendix}.

\subsubsection{Removing contamination first}


\begin{figure*}
    \centering
    \includegraphics[width=1\textwidth]{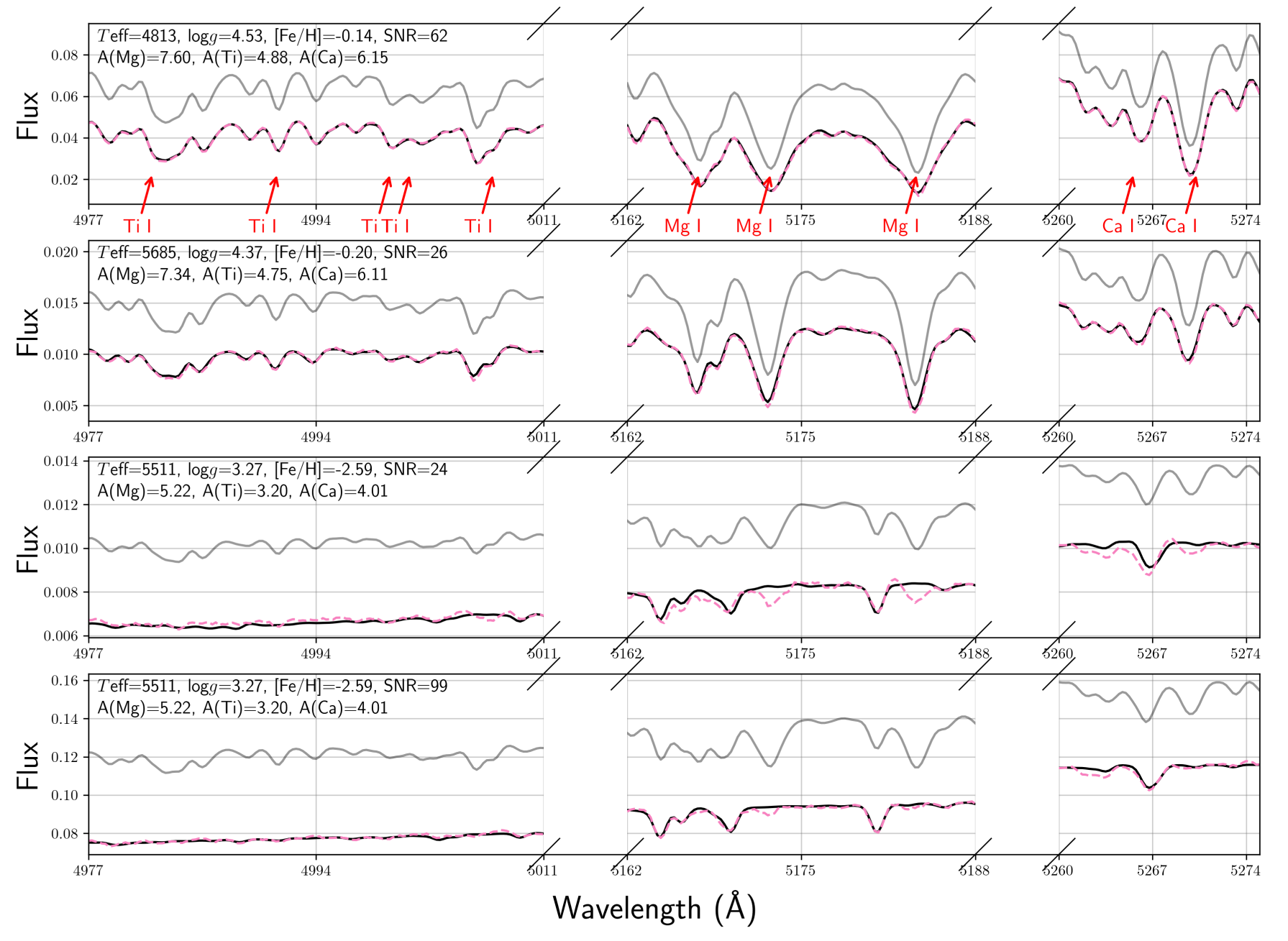}
    \caption{A Wave U-Net architecture was tasked with inferring (pink dashed line) the pure stellar spectrum (black solid line) from spectra contaminated by $\sim$40\% with a solar spectrum (grey solid line). Shown here are several examples, in the Mg I \textit{b} triplet region and regions with strong Ti I and Ca I absorption, of the inferred \texttt{WEAVE-GES-LR} stellar spectra with their contamination removed. The last two examples are of the same low-metallicity star at different SNR, which highlights the difficulty Wave U-Net has with low-metallicity and low SNR spectra. Aside from that, the predictions closely match the ground truth, though some residuals can be seen in the deeper absorption features.}
    \label{fig:waveunet_example}
\end{figure*}

\begin{figure*}
    \centering
    \includegraphics[width=0.9\textwidth]{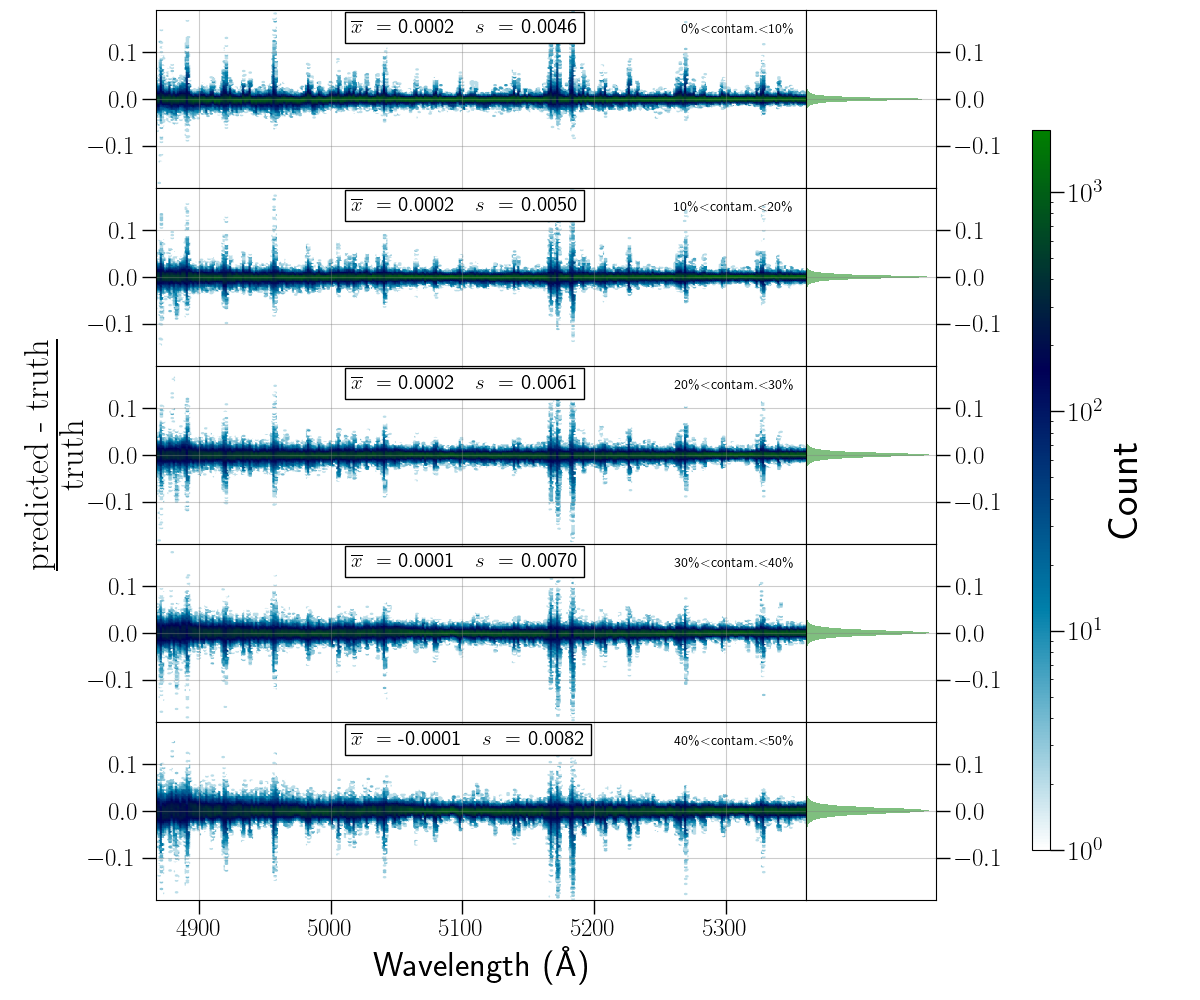}
    \caption{Wave U-Net was used to remove solar contamination from the test set of \texttt{WEAVE-GES-LR} stellar spectra, and shown here are the distributions of residuals between the predicted spectra and true spectra split into five groups of increasing levels of contamination. The bias, $\overline{x}$, and 1-sigma error, \textit{s}, were calculated for each group of spectra; the error in all cases is $<$1\% and gradually increases with the level of contamination. The residuals around strong absorption features (e.g. the Mg I \textit{b} triplet region around 5175 \AA) are more severe.}
    \label{fig:waveunet_resid}
\end{figure*}

\begin{figure*}
    \centering
    \includegraphics[width=0.95\textwidth]{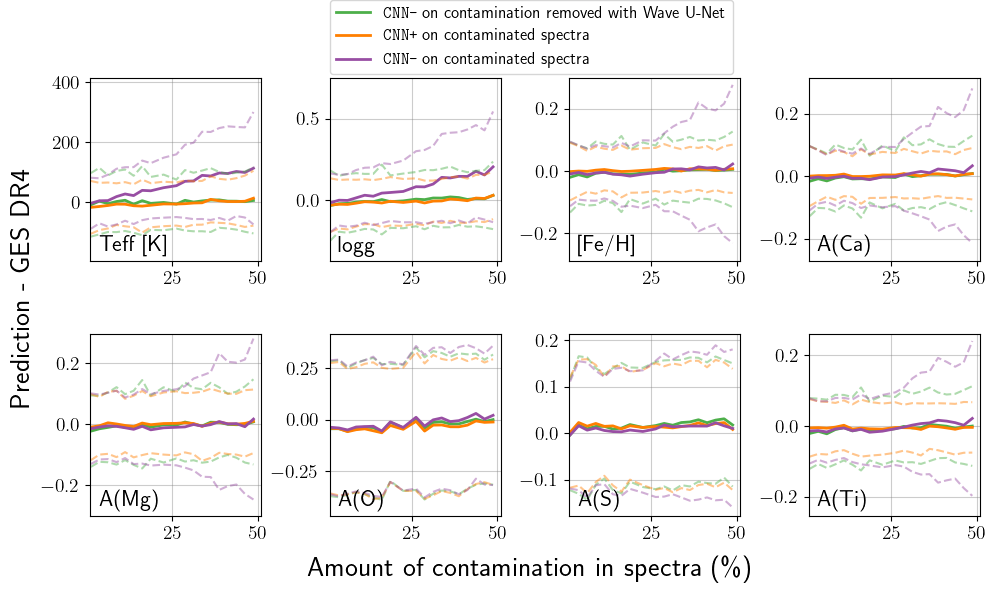}
    \caption{The residuals between stellar parameter and abundance predictions and GES catalog values, with the mean (solid lines) and standard deviation (dashed lines) of the residuals binned according to the amount of contamination in the original spectra. Two CNN models were separately trained on \texttt{WEAVE-GES-LR} spectra: one with a training set that included contamination (\texttt{CNN+}), and one that did not (\texttt{CNN-}). Each model was used to predict the stellar parameters $T$eff, logg, [Fe/H], and elemental abundances A(Ca), A(Mg), A(O), A(S), A(Ti) on a test set of \texttt{WEAVE-GES-LR} contaminated spectra. Additionally, \texttt{CNN-} was used for predictions on a test set of \texttt{WEAVE-GES-LR} spectra with their contamination removed by Wave U-Net. Overall better predictions were achieved with the \texttt{CNN+}. Predictions on the abundances O and S had large errors in all cases because of the lack of strong absorption features in the limited wavelength range studied.}
    \label{fig:stellar-param-resids-weaveLR}
\end{figure*}

\begin{figure*}
    \centering
    \includegraphics[width=0.95\textwidth]{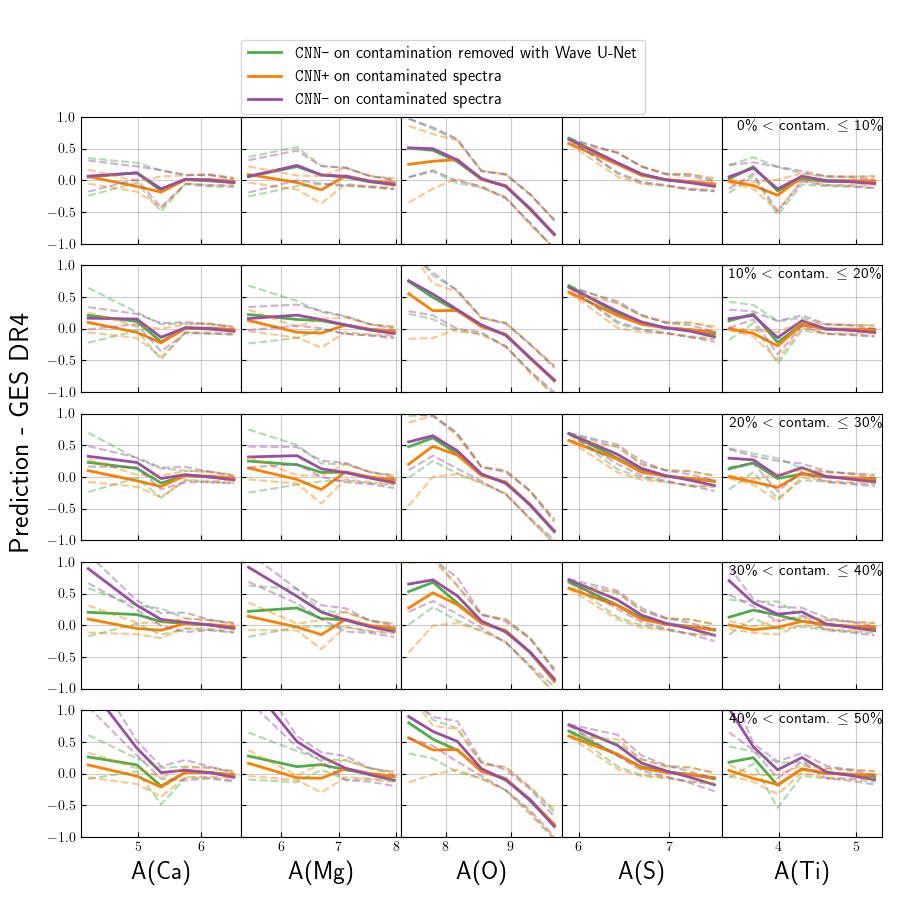}
    \caption{The residuals between stellar abundance predictions and GES catalog values. The mean (solid lines) and standard deviation (dashed lines) of the residuals were binned according to the abundance value, and each row corresponds to increasing levels of contamination in steps of 10\%. Two CNN models were separately trained on \texttt{WEAVE-GES-LR} spectra: one with a training set that included contamination (\texttt{CNN+}), and one that did not (\texttt{CNN-}). Each model was used to predict the stellar elemental abundances A(Ca), A(Mg), A(O), A(S), A(Ti) on a test set of \texttt{WEAVE-GES-LR} contaminated spectra. Additionally, \texttt{CNN-} was used for predictions on a test set of \texttt{WEAVE-GES-LR} spectra with their contamination removed by Wave U-Net. At higher abundance values and lower contamination levels, there is almost no difference between the three models. The differences become pronounced at lower abundance values and higher contamination levels, where \texttt{CNN+} can be seen to perform best.}
    \label{fig:abund-resids-weaveLR}
\end{figure*}


One method for dealing with the satellite contamination explored in this study was to attempt removing its signature from the observed spectra completely -- a process analogous to subtracting off a sky spectrum to isolate the pure stellar spectrum. Once the spectra are cleaned of their contamination, they can be sent to any downstream pipeline to estimate stellar parameters. As described in Section \ref{section: waveunet}, the Wave U-Net architecture was used for this task.

Figure \ref{fig:waveunet_example} shows several examples of Wave U-Net recovering \texttt{WEAVE-GES-LR} stellar spectra in regions with strong Mg, Ti, and Ca features with $\sim$40\% solar contamination: in most cases, the contamination was almost completely removed and the stellar spectra were recovered but with some more obvious errors around the deeper absorption features, especially for stars with low metallicity and low SNR. To see the overall performance of Wave U-Net the relative residuals of its predictions on the entire test set of \texttt{WEAVE-GES-LR} spectra were calculated, and the distributions are shown in Figure \ref{fig:waveunet_resid}. The 1-sigma errors were under 1\% across the entire test set, with a minimum of 0.46\% for the spectra with 0-10\% contamination, though it was confirmed that the model has larger residuals in the deeper absorption features.

With Wave U-Net operational, we obtained a test set of \texttt{WEAVE-GES-LR} spectra with their contamination removed to test the performance of \texttt{CNN-}. Figure \ref{fig:stellar-param-resids-weaveLR} shows the mean and standard deviation of residuals of \texttt{CNN-}'s stellar parameter predictions on the test set of \texttt{WEAVE-GES-LR} contaminated spectra before and after being processed by Wave U-Net. The residuals on the contaminated spectra are nearly identical to those on the Wave U-Net spectra in the low contamination regime, $\sim$2-3x higher in the high contamination regime, and significantly increase for the abundances around the 25\% contamination level. The residuals on the Wave U-Net spectra are relatively constant across most contamination levels, though a slight increase can be seen at the highest levels.

\subsubsection{Training a CNN on contaminated spectra}

Instead of identifying and throwing away a contaminated spectrum, or attempting to remove its contamination, one might be interested in extracting accurate information \textit{despite} the contamination. One way of accomplishing this task is again to use ML.

A major benefit of the feed-forward CNN architecture used in this study is that the spectra can be augmented with features (e.g. solar contamination) to be identified or, in this case, \textit{ignored}. By feeding artificially contaminated spectra through the NN and giving it the task of predicting stellar parameters of \textit{un}contaminated spectra, the NN learns to be robust to contamination in its predictions. 

To demonstrate the ability of a CNN to predict through contamination, the \texttt{CNN+} model was given a test set of contaminated spectra, and 
Figure \ref{fig:stellar-param-resids-weaveLR} shows the residuals of its predictions on the \texttt{WEAVE-GES-LR} spectra. The \texttt{CNN+} residuals are constant and overall lower than \texttt{CNN-}'s residuals on both the Wave U-Net spectra and contaminated spectra over the full range of contamination and across all parameters. Even with 0-2\% contamination levels, the residuals for $T_{\textrm{eff}}$ and log$g$ are slightly larger in \texttt{CNN-} than for \texttt{CNN+}. Note that in the wavelength range studied, there are limited oxygen and sulfur features and thus the residuals remain large and somewhat constant for both \texttt{CNN+} and \texttt{CNN-}. 

Figure \ref{fig:abund-resids-weaveLR} shows a similar comparison but the residuals were instead binned according to the stellar abundances, and split into 5 groups of increasing contamination. At low (0-10\%) contamination levels, the results are nearly identical. At lower abundance levels, it can be seen that Wave U-Net has difficulty with properly recovering the spectra even at lower (10-20\%) contamination levels, leading to worse precision than even the \texttt{CNN-} on contaminated spectra. The performance of the \texttt{CNN+} becomes especially apparent at higher contamination levels and low abundance levels, whereas all three models perform equally well at higher abundance levels.

\subsubsection{Removing contamination for downstream tasks}

\begin{figure*}
    \centering
    \includegraphics[width=0.95\textwidth]{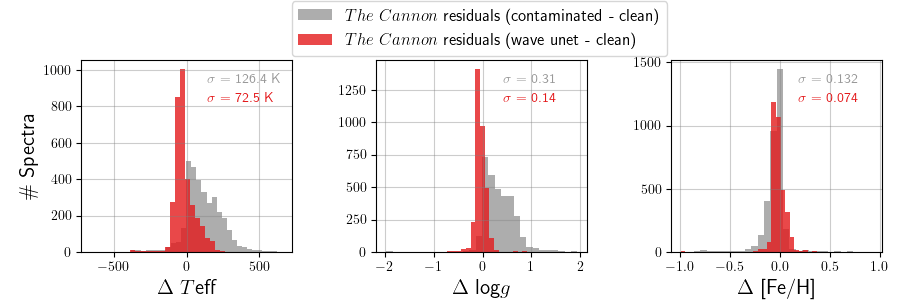}
    \caption{$The~Cannon$ was used to predict the stellar parameters $T$eff, log$g$, and [Fe/H] on three test sets of \texttt{WEAVE-GES-LR} spectra: contaminated (by $<$50\%) spectra, clean spectra, and spectra with contamination removed by Wave U-Net. The residuals between the predictions on the contaminated spectra and clean spectra, as well as the residuals between the predictions on Wave U-Net spectra and clean spectra, were computed. The histograms show the improvements in stellar parameter estimates when Wave U-Net is used to remove the contamination, confirming that Wave U-Net can be used in conjunction with other stellar parameter estimation pipelines.}
    \label{fig:thecannon-residuals-weaveHR}
\end{figure*}

To show the utility of removing contamination outside the scope of neural networks and our pipeline, $The~Cannon$ \citep{ness2015cannon} was used to predict the stellar parameters $T_{\textrm{eff}}$, log$g$, and [Fe/H] on contaminated, clean, and Wave U-Net \texttt{WEAVE-GES-LR} spectra. Figure \ref{fig:thecannon-residuals-weaveHR} shows the substantial reduction in prediction residuals -- the standard deviation reduces by a factor of $\sim$2 -- when the contaminated spectra are cleaned by Wave U-Net first.

\section{Discussion}

\subsection{Limitations}

\subsubsection{Low levels of contamination}

The results from our study show that the majority of low-contamination ($<$5\% contamination) spectra are not detected with the presented framework. Since low contamination will likely be the most prevalent level of contamination (especially in the \texttt{WEAVE-HR} spectra), this could mean that many contaminated spectra would pass through our current pipeline undetected. This creates a problem if the goal is simply to flag the contaminated spectra for removal or further processing, but there isn't necessarily a problem for stellar parameter and abundance predictions. Indeed, Figures \ref{fig:stellar-param-resids-weaveLR} and \ref{fig:stellar-param-resids-weaveHR} show that the errors on abundance predictions due to satellite contamination are insignificant compared to typical spectroscopic uncertainties \citep[e.g.,][]{recio2022gaia, accetta2022seventeenth, steinmetz2020sixth, aguado2019pristine, smiljanic2016gaia} when contamination is $\lesssim$20\% for \texttt{WEAVE-GES-LR} and $\lesssim$10\% for \texttt{WEAVE-GES-HR}.

For temperature and surface gravity predictions, however, Figures \ref{fig:stellar-param-resids-weaveLR} and \ref{fig:stellar-param-resids-weaveHR} show that even low levels of contamination can introduce noticeable errors. When Wave U-Net is used to remove the low-level contamination, the resulting predictions are at best marginally better and sometimes marginally worse than those on contaminated spectra; removing very low levels of contamination is difficult for a Wave U-Net. By including contamination as an augmentation to a training set instead of attempting to remove it, such that a CNN learns to bypass (or ignore) the contamination when making stellar parameter predictions, the errors on temperature and surface gravity estimates are reduced. Contamination as an augmentation therefore appears to be the best strategy for dealing with spectra containing low (and high) levels of contamination, though Wave U-Net is only marginally worse.

\subsubsection{Realistic training set}

In this study it was assumed the spectrum reflected by the satellites would be an unmodified solar spectrum, but in reality the satellites from each company will have different albedos and thus reflect a slightly modified solar spectrum. The results in this paper should therefore be interpreted as a best-case scenario of knowing the characteristics of the contamination completely. There are two options to include more realistic contamination information in the pipeline: (1) use the published materials, if available, from each company regarding the reflection properties of their satellites and include these as a wavelength-dependent data augmentation in the training set, and/or (2) the telescope could intentionally observe stars that either have previously had confirmed satellite contamination in their spectra or are predicted to have a satellite crossing their path, to collect real examples for the training set. Ideally both options would be pursued to curate a realistic training set, however   
it is not possible -- with current data that each company supplies -- to predict satellite positions with enough accuracy to know whether they have contaminated or will contaminate a star spectrum at a given time in a given place. We do expect that there will be some studies (both from industry and from astronomers) aimed at characterizing the reflectivity of various classes of satellites. 

\subsubsection{Solar twins}

\begin{figure*}
    \centering
    \includegraphics[width=0.95\textwidth]{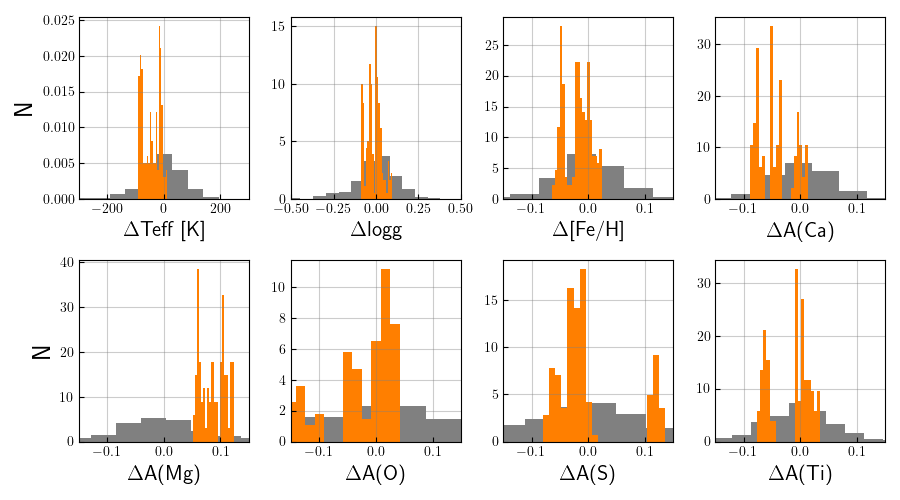}
    \caption{A set of \texttt{WEAVE-GES-LR} solar twins (defined as having $T_{\textrm{eff}}$ within 100\,K, $\textrm{log}g$ within 0.1\,dex, and [Fe/H] within 0.04\,dex of the solar values) was collected and \texttt{CNN+} was used to predict their stellar parameters and abundances and compare to GES catalog values. Shown here are the normalized distributions of residuals (predicted minus catalog values) on the solar twins (orange) and the rest of the test set (grey). The residuals are mostly compatible with the overall distribution, but there is a consistent over-prediction of magnesium abundance and a small under-prediction of temperature. Note: only five solar twins were found in the test set, each with multiple instances of contamination and all with |$v_\textrm{rad}|<15$km/s.}
    \label{fig:solartwinsparampredictions}
\end{figure*}

A star with solar-like stellar parameters presents an added difficulty for a NN since the contaminating and true spectra are similar and thus harder to disentangle. This was indeed the case for the identification task, where the CNN had a lower recall around solar-like surface gravity and metallicity. For the task of predicting stellar parameters and abundances, however, \texttt{CNN+} can help overcome this limitation. A set of five solar twins -- defined as having $T_{\textrm{eff}}$ within 100\,K, $\textrm{log}g$ within 0.1\,dex, and [Fe/H] within 0.04\,dex of the solar values (a slightly narrower definition than used by \citealt{ramirez2009accurate})  -- was used to check if their predicted parameters were significantly affected by contamination. Figure \ref{fig:solartwinsparampredictions} shows that the distributions of the stellar parameters and abundances for the solar twins are compatible with the overall distributions from the test set with two exceptions: (1) a consistent and significant over-prediction of 0.1\,dex for the magnesium abundances, and (2) an insignificant under-prediction of $<$100\,K for temperature. We suggest that even low levels of satellite contamination can impact magnesium abundance predictions in solar twins. With a sample size of five, however, the biases we see could also come from an under-representation of solar twins in the test set; a larger test set is required to make statistically significant conclusions.

\subsection{Further applications}

Another way to view the results in this paper is the application of a CNN to bright sky subtraction.  Typically high resolution spectroscopy is carried out at observatories during "bright time", when the face of the Moon is $>50\%$ illuminated, leading to stellar spectra with higher levels of solar contamination. There is nothing fundamentally different about a satellite reflecting solar light and the moon reflecting solar light (aside from wavelength-dependent reflectance properties) so the methods outlined in this paper could be directly applied to bright time stellar spectra. 

A similar method could also be used to detect whether a measured spectrum contains light from a binary stellar system. A training data set could be created in which stellar spectra are "contaminated" with a library of artificial binary counterparts; a CNN could be trained to detect the binary systems while a U-Net could be trained to separate the two spectra. One would need to carefully consider realistic values for the relative differences in radial velocity, temperature, surface gravity, abundances, and brightness between the artificially created binary stellar spectra.


\section{Conclusions}
\label{section:conclusions}

The detection and mitigation of contaminated spectra are important tasks in astronomical research, as contamination can lead to inaccurate conclusions and bias in data analysis. In this study, we applied a convolutional neural network (CNN) for both detection and stellar parameter prediction (specifically stellar parameters $T_{\textrm{eff}}$, log$g$, and [Fe/H], and the abundances A(Ca), A(Mg), A(O), A(S), and A(Ti)), and a Wave U-Net for removal of satellite contamination in a WEAVE-like spectral survey. We determined the limitations and feasibility of a machine learning-based pipeline for these tasks. 

In a test set of 10,000 artificially contaminated \texttt{WEAVE-GES-HR} spectra, the decision threshold of a CNN was adjusted such that a detection precision of 96\% and a detection recall of 67\% were achieved, which means that while the model misses 33\% of the total contaminated sources (mostly those with $<$10\% contamination), it can confidently identify the contamination it does detect. The results on \texttt{WEAVE-GES-LR} spectra were slightly poorer, with a precision of 80\% at the same recall level. It was also found that a contaminated spectrum is harder to identify if it has solar-like surface gravity, solar-like metallicity, or low radial velocity. 

A Wave U-Net model was trained to recover clean spectra from contaminated spectra and was able to recover them with 1-$\sigma$ errors of 0.46\% in the low ($<10\%$) contamination regime and $<$1\% in the high (40-50\%) contamination regime. Two CNNs were separately trained on contaminated (\texttt{CNN+}) and clean (\texttt{CNN-}) spectra, and it was shown that the accuracy in predictions, when compared to \texttt{CNN-} predictions on contaminated spectra, was improved by up to a factor of 2-3 when using \texttt{CNN-} on the Wave U-Net spectra and when using \texttt{CNN+} on contaminated spectra, with overall better results from the \texttt{CNN+}. Including an expected contamination as an artificially augmented training set therefore appears to be the best strategy for mitigating the effects of contamination in our pipeline. 




\section*{Acknowledgements}

We thank Federico Sestito, John Pazder, and Ted Grosson for the helpful comments and discussions about bright time sky subtraction, satellite coatings, and LSST satellite contamination. We also thank the anonymous referee for comments which helped strengthen and clarify our results.
SB and KAV thank the Natural Sciences and Engineering
Research Council for funding through the Discovery Grants
program and the CREATE program in New Technologies
for Canadian Observatories. SL acknowledges the support by PRIN INAF 2019 grant ObFu 1.05.01.85.14 (“Building up the halo: chemo-dynamical tagging in the age of large surveys”, PI. S. Lucatello).

\textit{Software}: \texttt{astropy} \citep{robitaille2013astropy}, \texttt{matplotlib} \citep{hunter2007matplotlib},  \texttt{numpy} \citep{harris2020array}, \texttt{pytorch} \citep{paszke2019pytorch},  \texttt{scipy} \citep{virtanen2020scipy}, \texttt{seaborn} \citep{waskom2021seaborn}.


\section*{Code availability}

All code for this project can be found at the project's Github repository: \url{https://github.com/astroai/StarUnLink}.

\section*{Data availability}

All data underlying this article will be shared on reasonable request to the corresponding author. 



\bibliographystyle{mnras}
\bibliography{bibliography}


\newpage
\appendix
\section{Results on high-resolution spectra}
\label{appendix}

\begin{figure*}
    \centering
    \includegraphics[width=0.9\textwidth]{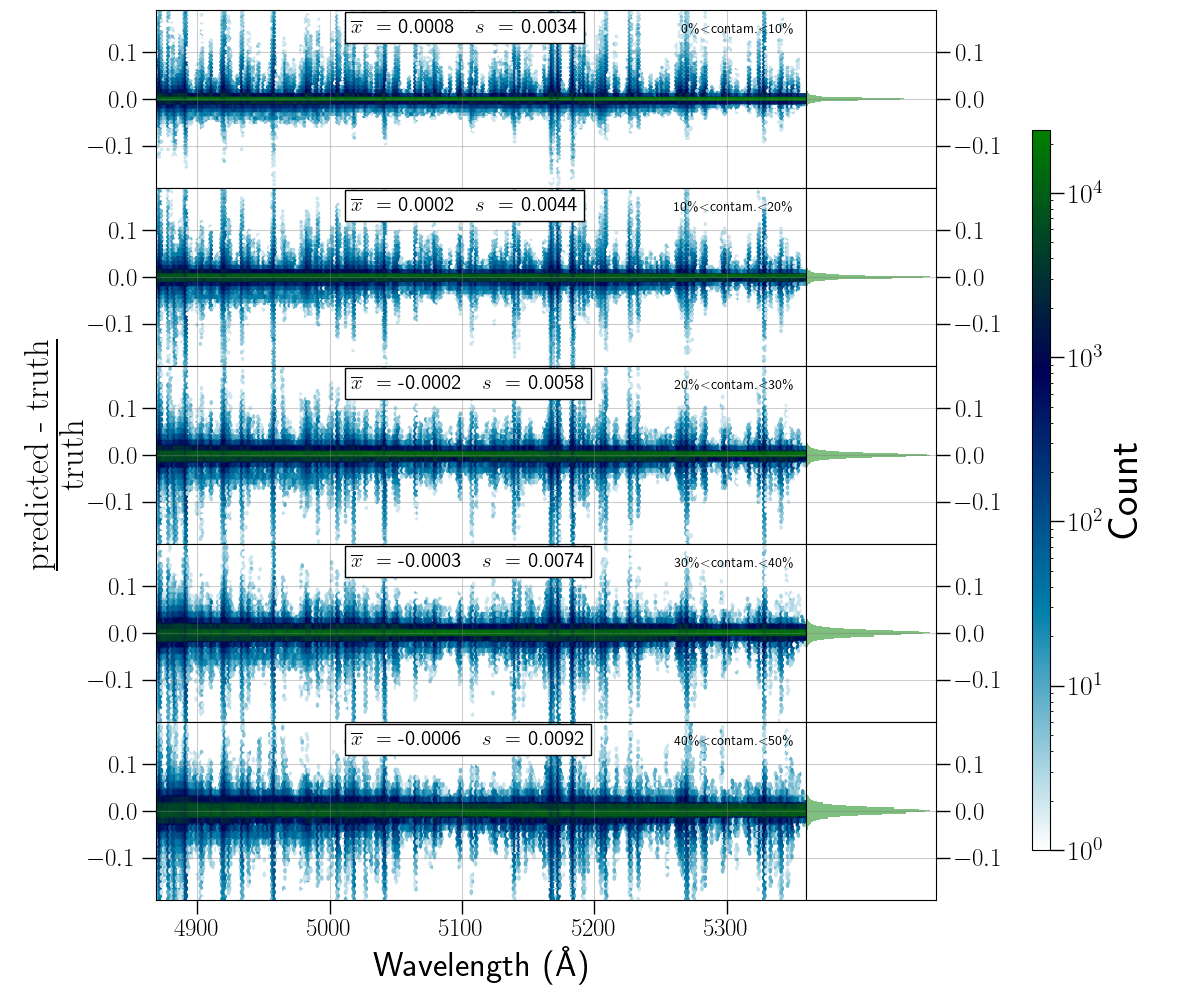}
    \caption{Wave U-Net was used to remove solar contamination from the test set of \texttt{WEAVE-GES-HR} stellar spectra, and shown here are the distributions of residuals between the predicted spectra and true spectra split into five groups of increasing levels of contamination. The bias, $\overline{x}$, and 1-sigma error, \textit{s}, were calculated for each group of spectra; the error in all cases is $<$1\% and gradually increases with the level of contamination. The residuals around strong absorption features (e.g. the Mg I \textit{b} triplet region around 5175 \AA) are more severe.}
    \label{fig:waveunet_resid-HR}
\end{figure*}

\begin{figure*}
    \centering
    \includegraphics[width=0.95\textwidth]{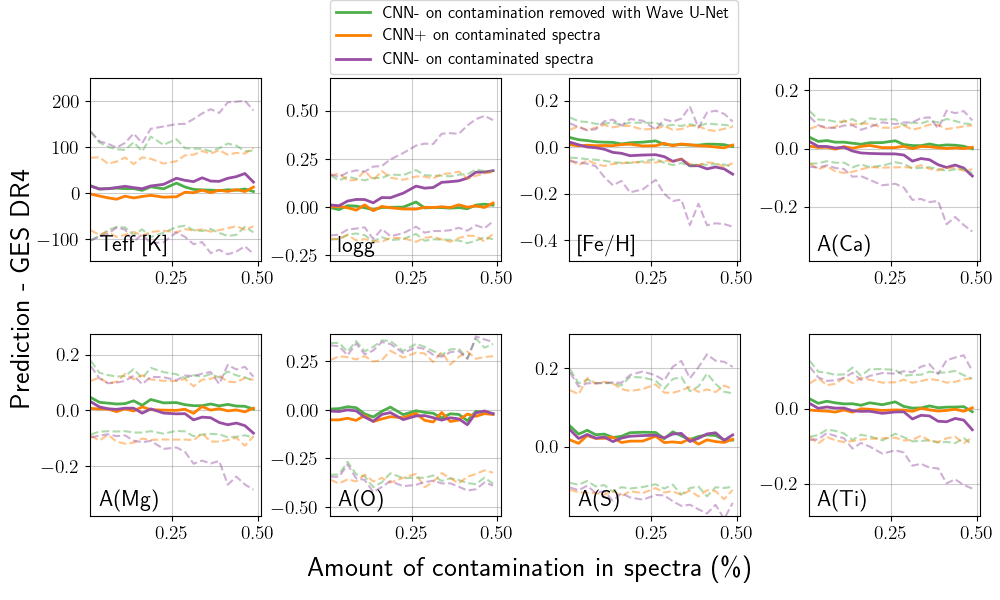}
    \caption{The residuals between stellar parameter and abundance predictions and GES catalog values, with the mean (solid lines) and standard deviation (dashed lines) of the residuals binned according to the amount of contamination in the original spectra. Two CNN models were separately trained on \texttt{WEAVE-GES-HR} spectra: one with a training set that included contamination (\texttt{CNN+}), and one that did not (\texttt{CNN-}). Each model was used to predict the stellar parameters $T$eff, logg, [Fe/H], and elemental abundances A(Ca), A(Mg), A(O), A(S), A(Ti) on a test set of \texttt{WEAVE-GES-HR} contaminated spectra. Additionally, \texttt{CNN-} was used for predictions on a test set of \texttt{WEAVE-GES-HR} spectra with their contamination removed by Wave U-Net. Overall better predictions were achieved with the \texttt{CNN+}. Predictions on the abundances O and S had large errors in all cases because of the lack of strong absorption features in the limited wavelength range studied.}
    \label{fig:stellar-param-resids-weaveHR}
\end{figure*}

The stellar parameters $T_{\textrm{eff}}$, log$g$, and [Fe/H], along with the absolute abundances A(Ca), A(Mg), A(O), A(S), and A(Ti), were predicted with a CNN. For the \texttt{WEAVE-GES-HR} sample, two models were trained: one on 35,000 spectra \textit{with} contamination between 0\% and 50\% (\texttt{CNN+}) and one on 35,000 spectra \textit{without} contamination (\texttt{CNN-}). Additionally, a Wave U-Net was used to remove the contamination from the \texttt{WEAVE-GES-HR} sample. Figure \ref{fig:waveunet_resid-HR} shows the residuals in the reconstructed Wave U-Net spectra, and Figure \ref{fig:stellar-param-resids-weaveHR} shows the residuals in the stellar parameter and abundance predictions of \texttt{CNN+} and \texttt{CNN-}.

\bsp	
\label{lastpage}
\end{document}